\documentclass[letterpaper,aps,prb,twocolumn,floatfix,amsfonts,amssymb,notitlepage]{revtex4-1}
\pdfpageattr {/Group << /S /Transparency /I true /CS /DeviceRGB>>}

\usepackage{amsmath}
\usepackage{amsthm}
\usepackage{amssymb}
\usepackage{graphicx}
\usepackage{tabularx}
\usepackage{color}
\usepackage{txfonts}
\usepackage{subfigure}
\usepackage{sidecap}

\newcommand{\cre}[2]{#1_{#2}^\dagger} 
\newcommand{\ann}[2]{#1_{#2}^{\phantom{\dagger}}} 
\newcommand{\crev}[2]{\vec{#1}_{#2}^\dagger} 
\newcommand{\annv}[2]{\vec{#1}_{#2}^{\phantom{\dagger}}} 
\def\kpar{\textbf{k}_{\parallel}}

\begin{document}
\title{Signatures of Weyl semimetals in quasiparticle interference}
\author{Andrew K. Mitchell}
\author{Lars Fritz}
\affiliation{Institute for Theoretical Physics, Utrecht University, 3584 CE Utrecht, The Netherlands}

%%%%%%%%%%%%%%%%%%%%%%%%%%%%%%%%%%%%%%%%%%%%%%%%%%%%%%%%%%%%%%%%%%%%%%%
%%%%%%%%%%%%%%%%%%%%%%%%%%%%%%%%%%%%%%%%%%%%%%%%%%%%%%%%%%%%%%%%%%%%%%%

\begin{abstract}
Impurities act as \emph{in situ} probes of nontrivial electronic structure, causing real-space modulations in the density of states detected by scanning tunneling spectroscopy on the sample surface. We show that distinctive topological features of Weyl semimetals can be revealed in the Fourier transform of this map, interpreted in terms of quasiparticle interference (QPI). 
We develop an exact Green's function formalism and apply it to generalized models of Weyl semimetals with an explicit surface. 
The type of perturbation lifting the Dirac node degeneracy to produce the 3D bulk Weyl phase determines the specific QPI signatures appearing on the surface. 
QPI Fermi arcs may or may not appear, depending on the relative surface orientation and quantum interference effects. Line nodes give rise to tube projections of width controlled by the bias voltage. 
We consider the effect of crystal warping, distinguishing dispersive arc-like features from true Fermi arcs. Finally we demonstrate that the commonly used joint-density-of-states approach fails qualitatively, and cannot describe QPI extinction. 
\end{abstract}
\date{\today}
\maketitle

%%%%%%%%%%%%%%%%%%%%%%%%%%%%%%%%%%%%%%%%%%%%%%%%%%%%%%%%%%%%%%%%%%%%%%%
%%%%%%%%%%%%%%%%%%%%%%%%%%%%%%%%%%%%%%%%%%%%%%%%%%%%%%%%%%%%%%%%%%%%%%%

\section{Introduction}
\label{sec:intro}

Weyl semimetals (WSMs) are topologically nontrivial states of quantum matter characterized by the existence of three-dimensional chiral Dirac nodes.\cite{wan2011topological,balents2011viewpoint,weng2015weyl,huang2015weyl,turner2013beyond,Lv2015Experimental,xu2015discovery,xu2015discovery2,yang2015weyl,lv2015observation,xu2015observation,shekhar2015extremely} Unlike 3D Dirac semimetals, Weyl nodes of different chirality are nondegenterate, and separated in momentum space. As such, they act as sources and sinks of Berry curvature.\cite{wan2011topological} WSMs have been likened\cite{yang2015weyl} to ``3D graphene'', but also exhibit a range of fascinating properties not observed in graphene or other topological semimetallic systems -- such as the chiral anomaly in quantum transport\cite{zyuzin2012topological,liu2013chiral,zhang2015observation,PhysRevB.92.125141} and the appearance of \emph{open} surface Fermi arcs in photoemission measurements.\cite{Lv2015Experimental,xu2015discovery,xu2015discovery2,yang2015weyl,lv2015observation,xu2015observation} The latter are the result of topologically-protected chiral states connecting bulk Weyl nodes of opposite chirality, projected onto the surface. Information about the bulk topology can therefore be obtained from the surface-projected Fermi arcs. 
Such topological properties are predicted to be robust to weak perturbations, including disorder from dilute impurities,\cite{bera2015dirty} since the Weyl nodes can only be annihilated in pairs of opposite chirality.\cite{wan2011topological} 

A Weyl phase is realized by splitting degenerate nodes in a 3D Dirac semimetal in momentum and/or energy space, and necessarily involves breaking either inversion or time-reversal symmetry. This can be done in a number of different ways;\cite{mitchell2015kondo} the microscopic details in real WSM systems can translate to different types of perturbations in the pristine low-energy Dirac theory. Depending on the particular perturbation arising in a given WSM, and the relative orientation of the material surface to the inter-node vector, it is to be expected that a \emph{range} of distinctive features could appear in surface measurements. Such surface probes could therefore be used to identify and fingerprint properties of the bulk.

A family of WSMs was predicted from band structure calculations in the monopnictide class\cite{weng2015weyl,huang2015weyl} -- and very recently Weyl fermion states have been discovered experimentally in the noncentrosymmetric (but time-reversal invariant) materials TaAs, NbAs, TaP and NbP.\cite{Lv2015Experimental,xu2015discovery,xu2015discovery2,yang2015weyl,lv2015observation,xu2015observation,shekhar2015extremely} In particular, surface Fermi arcs have been observed in angle-resolved photoemission spectroscopy (ARPES) experiments on these systems.

Another powerful technique for probing surface states is scanning tunneling spectroscopy (STS).\cite{STS,hoffman2002imaging} Impurities or potential defects in materials produce real-space modulations (Friedel oscillations\cite{hosur2012friedel}) in the surface density of states, and can be detected by STS. In the case of magnetic impurities (such as transition metal adatoms), Kondo effects\cite{hewson1997kondo} can lead to different electronic scattering mechanisms and distinctive spectroscopic signatures.\cite{ternes2009spectroscopic,mitchell2015kondo} For systems with either static Born-type impurities\cite{RevModPhys.78.373} or dynamic Kondo impurities,\cite{derry2015quasiparticle,mitchell2015multiple} the Fourier transform of the real-space STS density map (FT-STS) can be interpreted in terms of quasiparticle interference (QPI). 
The impurity-induced scattering of quasiparticles is both energy and momentum dependent. At a given energy
(set by the bias voltage in an STS experiment\cite{tersoff1983theory}), the quantum interference between different scattering processes produces the patterns observed in QPI. The scattering, and therefore the QPI, is \emph{entirely characteristic} of the host material -- meaning that impurities act as sensitive \emph{in situ} probes of its electronic structure. 

Important insights into various materials have been gained from QPI -- including systems with nontrivial topology, such as 3D topological insulators.\cite{roushan2009topological,kuroda2010hexagonally,mitchell2013kondo} Indeed, Fermi arcs in the cuprate superconductors have also been extensively investigated with QPI\cite{hoffman2002imaging,hoffman2002four,Lee28082009} (albeit that their physical origin is very different from that of WSMs). Experimental QPI patterns for the 3D Dirac semimetal Cd$_3$As$_2$ were also obtained in Ref.~\onlinecite{jeon2014landau}, complementing ARPES results. Very recently, the first atomic-scale visualization of a WSM surface was obtained by STS for the material NbP in Ref.~\onlinecite{zheng2015atomic}. The detailed study of topological Weyl systems, using QPI as a sensitive surface probe, is now a possibility. 

In this paper, we examine theoretically the different types of topological signature that can appear in QPI for Weyl systems. For example, it is generally expected that distinctive features, such as the Fermi arcs observed in ARPES, could also be found in QPI.\cite{sau2012topologically,chang2015quasi} We find that intense inter-arc scattering indeed produces QPI Fermi arcs, although intra-arc scattering can be comparatively weak due to quantum interference effects. 
We present a generalized and exact formulation for calculation of QPI in terms of Green's functions and the scattering t-matrix for topological WSM models with an \emph{explicit surface}, in the presence of one or more impurities. Here we do not attempt to make realistic material-specific predictions, but rather we focus on generic effective models whose exact solutions enable us to establish how particular types of bulk structures in Weyl systems can be characterized in terms of their distinctive QPI patterns. We show, however, that lattice effects such as  crystal warping can be simply incorporated in a phenomenological way, so that contact can be made with future experiments.

We note that the popular and appealingly simple `joint density of states' (JDOS) interpretation of QPI\cite{hoffman2002imaging,sau2012topologically,chang2015quasi} is in fact only applicable in the simplest one-band case with a single static impurity.\cite{derry2015quasiparticle} The JDOS can yield qualitatively incorrect results for Weyl systems because it neglects quantum interference effects and does not account for the matrix structure of the scattering problem in spin-space (the JDOS approach is known to fail e.g.\ for graphene\cite{simon2011fourier} for similar reasons). The full (complex, dynamical) Green's functions must therefore be used, preserving the matrix structure inherent to topological multiband systems. 

In Sec.~\ref{sec:model}, we specify the basic degenerate 3D Dirac theory, and discuss briefly the effect of the different perturbations, classified exhaustively in Ref.~\onlinecite{mitchell2015kondo}, which separate the nodes in momentum space to produce a Weyl phase. We then employ a semi-infinite 3D model of coupled layers with an explicit surface, which yields the desired degenerate Dirac theory at low energies. We identify representative physical perturbations following Refs.~\onlinecite{burkov2011weyl,halasz2012time} that produce either TR-broken or TR-invariant Weyl phases, and also line-node semimetals.

Based on a matrix generalization of the equations of motion method,\cite{zubarev1960double} exact analytic expressions for the surface Green's functions of the WSM models are found in Sec.~\ref{sec:GF}. The t-matrix, describing scattering from one or more impurities is formulated, and used to obtain the full QPI in Sec.~\ref{sec:QPI}

In Sec.~\ref{sec:Dirac} we consider explicitly QPI signatures of Dirac cones in WSMs, with crystal warping effects explored in Sec.~\ref{sec:warp}. QPI Fermi arcs are discussed in Sec.~\ref{sec:FA} for the IS broken but TR-invariant case relevant to the monopnictide Weyl materials. QPI signatures of more exotic line-node WSMs are examined in Sec.~\ref{sec:LN}.

The failure of the JDOS approach is discussed in Sec.~\ref{sec:JDOS}.

%%%%%%%%%%%%%%%%%%%%%%%%%%%%%%%%%%%%%%%%%%%%%%%%%%%%%%%%%%%%%%%%%%%%%%%
%%%%%%%%%%%%%%%%%%%%%%%%%%%%%%%%%%%%%%%%%%%%%%%%%%%%%%%%%%%%%%%%%%%%%%%

\section{Models for Weyl Semimetals}
\label{sec:model}

We first consider the minimal low-energy Bloch theory for a pristine 3D Dirac semimetal,
\begin{eqnarray}\label{eq:DiracSM}
{\hat{\mathcal{H}}}_{\rm{D}} ({\bf{k}})=v_F \; {\hat{\tau}}_z  \otimes {\bf{k}}\cdot {\hat{{\vec{\sigma}}}}\;,
\end{eqnarray}
where $\hat{\sigma}$ and $\hat{\tau}$ are Pauli matrices acting respectively in spin and orbital space, and $v_F$ is the effective Fermi velocity. 
This model possesses both time-reversal symmetry (TRS) and inversion symmetry (IS) --- meaning  
$\mathcal{T}{\hat{\mathcal{H}}}_{\rm{D}}({\bf{k}})\mathcal{T}^{-1}={\hat{\mathcal{H}}}_{\rm{D}}(-{\bf{k}})$ in terms of the time-reversal operator $\mathcal{T}=\hat{\tau}_0 \otimes (i \hat{\sigma}_y)K$ (with complex conjugation denoted by $K$); and $\mathcal{P}{\hat{\mathcal{H}}}_{\rm{D}}({\bf{k}})\mathcal{P}^{-1}={\hat{\mathcal{H}}}_{\rm{D}}(-{\bf{k}})$, in terms of the inversion operator $\mathcal{P}=\hat{\tau}_x \otimes \hat{\sigma}_0$.

As discussed recently in Ref.~\onlinecite{mitchell2015kondo}, perturbations to Eq.~\ref{eq:DiracSM} that yield a WSM can be classified according to the symmetries they break. To leading order at low energies, perturbations must be of the generic form $\delta\hat{H}=\left ( \vec{a} \cdot \hat{\vec{\tau}} +a_0\hat{\tau}_0 \right )\otimes  \left (\vec{b}\cdot \hat{\vec{\sigma}} + b_0 \hat{\sigma}_0 \right)$. 
In its most general form, the parameters $\vec{a}$, $a_0$, $\vec{b}$, and $b_0$ can also be momentum-dependent, allowing e.g.\ for additional crystal warping effects of the underlying lattice, or the cone tilting/tipping phenomenon discussed recently in Refs.~\onlinecite{BergholtzBrouwer,BernevigTroyer}.
Any microscopic model describing Dirac/Weyl systems must therefore map onto ${\hat{\mathcal{H}}}_{\rm{D}} ({\bf{k}})+\delta\hat{H}$ at low energies. 

For simplicity in the following, we take $\vec{a}$, $a_0$, $\vec{b}$, and $b_0$ to be pure constants, independent of momentum. Note that $\delta\hat{H}_{1} = a_0 \hat{\tau}_0 \otimes \vec{b} \cdot \hat{\vec{\sigma}}$ breaks TRS and splits the Weyl nodes in momentum space along $\vec{b}$, whereas $\delta\hat{H}_{2} = a_x \hat{\tau}_x \otimes \vec{b} \cdot \hat{\vec{\sigma}}$ and $\delta\hat{H}_{3} = a_y  \hat{\tau}_y \otimes \vec{b} \cdot \hat{\vec{\sigma}}$ produce line-nodes in the plane perpendicular to $\vec{b}$, with either IS or TRS. With sufficiently strong crystal warping, the rotational symmetry of line nodes is spoiled, and pairs of Weyl nodes appear instead (see also Sec.~\ref{sec:LN}).

The bulk 3D host is described by $H_{\text{WSM}}= \int \frac{d^3 k}{(2 \pi)^3} \Psi^\dagger ({\bf{k}}) [\hat{\mathcal{H}}_{\text{D}}(\textbf{k})+\delta\hat{H}]\Psi ({\bf{k}})$, in terms of the 4-component conduction electron operators $\Psi ({\bf{k}})$, living in $\tau$- and $\sigma$-space.

%%%%%%%%%%%%%%%%%%%%%%%%%%%%%%%%%%%%%%%%%%%%%%%%%%%%%%%%%%%%%%%%%%%%%%%

\subsection{Explicit surface formulation}
\label{sec:model_surface}

To describe surface spectroscopic signatures and QPI, we now employ a semi-infinite 3D model with an explicit 2D surface. Rather than resorting to a full tight-binding model, we instead use a compactified prescription\cite{burkov2011weyl} based on coupled layers (each a 2D Dirac theory), terminating at the surface, which we take to be perpendicular to $\hat{z}$. The pseudospin $\tau_z=\pm$ can refer to two different bands in each layer, with opposite Fermi velocity $\pm v_F$. A term in the dispersion like $\sin(d k_z)$ then leads to the desired linear dependence on $k_z$ in the bulk at low energies (hereafter, we take the lattice constant $d\equiv 1$).   The full Hamiltonian is given by,
\begin{equation}
\label{eq:H}
H_{\text{WSM}} = \int \frac{d^2 \kpar}{(2\pi)^2} ~ \hat{\mathcal{H}}_{\text{WSM}}(\kpar) \;,
\end{equation}
with
\begin{equation}
\label{eq:Hweyl}
\begin{split}
\hat{\mathcal{H}}_{\text{WSM}}(\kpar) = \sum_{j=0}^{\infty} \Bigg [ &\sum_{\tau_z=\pm} \Big ( \crev{c}{\kpar,j \tau_z} \hat{\mathcal{H}}_{\tau_z}(\kpar) \annv{c}{\kpar,j \tau_z}\Big ) \\
& + \Delta_T(\kpar)\Big(\crev{c}{\kpar,j -}\annv{c}{\kpar,j +} + \text{H.c.}\Big) \\
& + \Delta_N(\kpar)\Big(\crev{c}{\kpar,j -}\annv{c}{\kpar,(j+1) +} + \text{H.c.}\Big)\Bigg ] \;,
\end{split}
\end{equation}
where $\crev{c}{\kpar,j\tau_z}\equiv [\cre{c}{\kpar,j \tau_z \uparrow},\cre{c}{\kpar,j \tau_z \downarrow}]$ are two-component operators for conduction electrons in layer $j\ge 0$ with orbital pseudospin index $\tau_z=\pm$ and momentum $\kpar=(k_x,k_y)$ in the 2D plane parallel to the surface. As before, $\sigma=\uparrow/\downarrow$ labels the physical electron spin. Each layer is described by,
\begin{equation}
\label{eq:TIsurface}
\hat{\mathcal{H}}_{\pm}(\kpar) = \pm v_F \left ( \hat{\sigma}_x k_y-\hat{\sigma}_y k_x \right ) + \vec{m}\cdot \hat{\vec{\sigma}} \pm \delta(\kpar)\hat{\sigma}_0 \;.
\end{equation}
The terms proportional to $\vec{m}$ and $\delta(\kpar)$ break TRS and IS, respectively, allowing the Weyl phase to be realized.

This setup can also be thought of as a generalization of the topological insulator (TI) multilayer system discussed in Refs.~\onlinecite{burkov2011weyl,halasz2012time}. In that case, the pseudospin $\tau_z=\pm$ is to be understood as the upper/lower surfaces of a thin slice of a 3D TI. The heterostructure comprises alternating layers of the TI and ordinary-insulator spacer layers. $\Delta_T(\kpar)$ and $\Delta_N(\kpar)$ are then the tunneling amplitudes through the topological insulator layers and the normal insulator layers, respectively. The perturbation $\vec{m}$ would correspond to TRS-breaking net magnetization (due e.g.\ to ordered magnetic impurities); while $\delta(\kpar)$ is a staggered potential, breaking IS.

%%%%%%%%%%%%%%%%%%%%%%%%%%%%%%%%%%%%%%%%%%%%%%%%%%%%%%%%%%%%%%%%%%%%%%%

\subsection{Crystal warping effects}
\label{sec:crystal_warp}
For the purposes of this paper, we simply take Eqs.~\ref{eq:H}--\ref{eq:TIsurface} as a concrete compactified model for generic 3D WSMs, with a 2D surface at $j=0$ and $\tau_z=+$. We will use it to study the surface QPI signatures that might arise due to different types of topological bulk structure. However, to some extent, material-specific features and details of real WSMs can be reproduced phenomenologically in this model (at least at low energies) through the momentum dependence of $\Delta_T(\kpar)$, $\Delta_N(\kpar)$, and $\delta(\kpar)$. In real materials, the continuous rotational symmetry of the Dirac cones is reduced to discrete symmetries due to the underlying crystal lattice structure. For example, the cubic warping common in 3D TI surfaces can be encoded through,\cite{Fu_cubicwarp}
\begin{eqnarray}\label{eq:w3}
\delta(\kpar)=\delta + W_3[ (v_F k^{+})^3 + (v_F k^{-})^3 ]\;,
\end{eqnarray}
where $k^{\pm}=k_x\pm \text{i}k_y$ as usual. Similarly, properties of crystal structures with two- and four-fold symmetry could be approximated by the expansion,\cite{halasz2012time} 
\begin{equation}
\begin{split}
\label{eq:delta_kpar}
\Delta_{T,N}(\kpar)=\Delta^{(0)}_{T,N} &+ \Delta^{(2)}_{T,N}\left [ (v_F k^{+})^2 + (v_F k^{-})^2 \right ] \\ &+ \Delta^{(4)}_{T,N}\left [ (v_F k^{+})^4 + (v_F k^{-})^4 \right ] + ... \;.
\end{split}
\end{equation}
In particular, note that the monopnictide Weyl materials\cite{Lv2015Experimental,xu2015discovery} have the four-fold point group symmetry $C_{4v}$. 
The effect of crystal warping on the QPI is considered in Sec.~\ref{sec:warp} (see also Fig.~\ref{fig:warp}).

%%%%%%%%%%%%%%%%%%%%%%%%%%%%%%%%%%%%%%%%%%%%%%%%%%%%%%%%%%%%%%%%%%%%%%%

\subsection{Weyl phases}
\label{sec:weyl_pert}
For $|\vec{m}|=0$ and $\delta(\kpar)=0$, Eqs.~\ref{eq:H}--\ref{eq:TIsurface} describe a degenerate 3D Dirac semimetal in the bulk (i.e. away from the surface, $j\gg 1$). Breaking TRS through finite $m_z$ (magnetization along the layer stacking direction) splits the nodes to realize a Weyl phase. Specifically, for constant $\Delta_{T,N}(\kpar)\equiv \Delta_{T,N}^{(0)}$ (and $\delta(\kpar)=0$), the Weyl nodes are split along the $\hat{z}$ direction when\cite{burkov2011weyl} $[\Delta_{N}^{(0)}-\Delta_{T}^{(0)}]^2<m_z^2<[\Delta_{N}^{(0)}+\Delta_{T}^{(0)}]^2$. This perturbation is equivalent to the generic case of $\delta{\hat{H}}_1$ discussed in relation to Eq.~\ref{eq:DiracSM} above. A key question considered in Sec.~\ref{sec:Dirac} and Fig.~\ref{fig:dirac} is: what signatures of this appear in QPI on the surface, which is perpendicular to $\hat{z}$?

Generalizing to arbitrary magnetization direction $\vec{m}$, leads to other, qualitatively distinct,  possibilities. For example, any finite $m_x$ leads to a \emph{line node} semimetallic state. The line node is a ring in the $yz$ plane perpendicular to the $\hat{x}$ magnetization direction. The surface-projected QPI signatures of this kind of topological structure are discussed in Sec.~\ref{sec:LN} and Fig.~\ref{fig:line}. This case is equivalent to the generic perturbation $\delta\hat{H}_2$.

Finally, we focus on the IS-breaking case (but with TRS intact). This is the situation relevant to the recently discovered monopnictide Weyl materials, where surface QPI Fermi arcs might be expected.
In our effective model, this symmetry-breaking is implemented via the finite perturbation $\delta(\kpar)\equiv \delta$. This case is equivalent to the generic perturbation $\delta\hat{H}_3$. In the context of the TI multilayer system in Ref.~\onlinecite{halasz2012time}, a (ring) line node was found in the $xy$ plane parallel to the surface. However, it was also noted that sufficiently strong crystal warping (due to e.g.\ finite $\Delta_{N}^{(2)}$) spoils the rotational symmetry around the ring node, leading instead to the formation of two pairs of Weyl nodes. In fact, we show that additional Weyl node pairs can be generated on further increasing the crystal warping strength. In Sec.~\ref{sec:FA} we study the QPI Fermi arcs on the surface, resulting from the appearance of these bulk Weyl node pairs --- see Fig.~\ref{fig:arc}.

In the following, we set the Fermi velocity $v_F\equiv 1$ for convenience.

%#################################
%#################################

\section{Green's functions and impurity scattering}
\label{sec:GF}

Electronic properties of the WSM material can be characterized by the Green functions $G^{j \tau_z \sigma}_{j'\tau_z'\sigma'}(\kpar,\omega) = \langle\langle \ann{c}{\kpar,j\tau_z\sigma} ; \cre{c}{\kpar,j'\tau_z'\sigma'} \rangle\rangle^0_{\omega}$, where $\langle\langle \hat{A} ; \hat{B} \rangle\rangle_{\omega}$ is the Fourier transform of the retarded correlator $-i\theta(t)\langle \{ \hat{A}(t) , \hat{B}(0) \} \rangle$, and the superscript `0' denotes the clean (impurity-free) system. We define a $2\times 2$ Green function matrix in spin-space, $[\boldsymbol{G}^{j\tau_z}_{j'\tau_z'}(\kpar,\omega)]_{\sigma,\sigma'} = G^{j\tau_z\sigma}_{j'\tau_z'\sigma'}(\kpar,\omega)$. In particular, we are interested in the surface Green functions with $j=0$ and $\tau_z=+$, since sites on the surface are probed by STS in experiment.  

A matrix formulation of standard equations of motion provides simple exact relations between Green functions in this system. The surface Green functions are expressed as,
\begin{equation}
\label{eq:eom_surf}
\boldsymbol{G}^{0+}_{0+}(\kpar,\omega) [(\omega+i0^{+})\boldsymbol{I}-\boldsymbol{H}_{+}(\kpar)] = \boldsymbol{I}  + \Delta_{T}(\kpar) \boldsymbol{G}^{0-}_{0+}(\kpar,\omega) \;.
\end{equation}
Away from the surface [$(j\tau_z) \ne (0+)$], we obtain generally,
\begin{equation}
\label{eq:eom_gen}
\begin{split}
\boldsymbol{G}^{j\tau_z}_{j'\tau_z'}(\kpar,\omega) &[(\omega+i0^{+})\boldsymbol{I}-\boldsymbol{H}_{\tau_z}(\kpar)]= \delta_{jj'}\delta_{\tau_z\tau_z'}\boldsymbol{I}  \\
&+ \Delta_{T}(\kpar) \boldsymbol{G}^{j\bar{\tau_z}}_{j'\tau_z'}(\kpar,\omega) + \Delta_{N}(\kpar) \boldsymbol{G}^{(j-\tau_z)\bar{\tau_z}}_{j'\tau_z'}(\kpar,\omega) \;,
\end{split}
\end{equation}
where $\bar{\tau_z}=-\tau_z$ and $(j-\tau_z)\equiv (j\mp 1)$ for $\tau_z=\pm$. The surface Green functions in the semi-infinite system are then given as solutions of a matrix quadratic equation, obtained by recursive application of Eq.~\ref{eq:eom_gen}:
\begin{equation}
\label{eq:GFsurf}
\begin{split}
&\boldsymbol{G}^{0+}_{0+}(\kpar,\omega) = \Big [ (\omega+i0^{+})\boldsymbol{I}-\boldsymbol{H}_{+}(\kpar)  \\
&-\Delta_{T}(\kpar)^2 \Big [ (\omega+i0^{+})\boldsymbol{I}-\boldsymbol{H}_{-}(\kpar) - \Delta_{N}(\kpar)^2 \boldsymbol{G}^{0+}_{0+}(\kpar,\omega) \Big ]^{-1} \Big ]^{-1} \;.
\end{split}
\end{equation}
Surface Green functions can always be evaluated efficiently numerically by iterating Eq.~\ref{eq:GFsurf}. In certain cases (as shown below), simple analytic expressions can be found in closed form from Eq.~\ref{eq:GFsurf}.

Although Eq.~\ref{eq:GFsurf} is specific to the present model, the matrix equations of motion technique used to obtain it is widely applicable.

%#################################
%#################################

\subsection{Bulk-Boundary relation for Green's functions}
\label{sec:bulkGF}

Bulk Green functions (with $j\rightarrow \infty$) can similarly be obtained from Eq.~\ref{eq:eom_gen}. However, the bulk can also be realized by coupling together the boundaries of two semi-infinite systems. This allows surface and bulk Green functions to be related:
\begin{equation}
\label{eq:GFbulk}
\Big [\boldsymbol{G}^{\infty +}_{ \infty +}(\kpar,\omega)\Big]^{-1} = \Big [\boldsymbol{G}^{0+}_{0+}(\kpar,\omega) \Big ]^{-1} - \Delta_{N}(\kpar)^2 \boldsymbol{G}^{0+}_{0+}(-\kpar,\omega) \;.
\end{equation}
The surface propagator for one of the two subsystems being joined can therefore be viewed as a self-energy correction to the other.

%#################################
%#################################

\subsection{Impurity problem and t-matrix}
\label{sec:imp}

The clean host Weyl semimetal described above will inevitably be subject to some degree of disorder in real samples. Impurities or defects on the surface cause potential scattering, whose effect can be probed by quasiparticle interference (QPI), as described in the following sections. The full Hamiltonian is then $H=H_{\text{WSM}}+\sum_{\gamma} H_{\text{imp}}^{\gamma}$, where each impurity $\gamma$ is located at real-space site $\textbf{r}_{\gamma}$. The dominant source of electronic scattering on the surface is from surface impurities. For simplicity, we therefore consider only surface impurities here ($j=0$ and $\tau_z=+$); although the generalization to include bulk impurities is straightforward. The impurities are taken to be local in space: 
\begin{equation}
\label{eq:Himp}
H_{\text{imp}}^{\gamma} = \crev{c}{\textbf{r}_{\gamma},0+} \boldsymbol{V}^{\gamma} \annv{c}{\textbf{r}_{\gamma},0+} \;,
\end{equation}
with $\crev{c}{\textbf{r}_{\gamma},j\tau_z} = [ \cre{c}{\textbf{r}_{\gamma},j\tau_z\uparrow}, \cre{c}{\textbf{r}_{\gamma},j\tau_z\downarrow} ] $, and where
\begin{equation}
\label{eq:rsop}
\cre{c}{\textbf{r}_{\gamma},j\tau_z\sigma} = \int \frac{d^2 \kpar}{2\pi} ~\text{e}^{\text{i}\textbf{r}_{\gamma}\cdot \kpar} ~\cre{c}{\kpar,j\tau_z\sigma}
\end{equation}
creates an electron localized at site $\textbf{r}_{\gamma}$ with spin $\sigma$ in orbital/surface $\tau_z$ of layer $j$.

The $2\times 2$ matrix $\boldsymbol{V}^{\gamma}$ describing the local potential due to impurity $\gamma$ has elements $V^{\gamma}_{\sigma\sigma'}$. The specific form of $\boldsymbol{V}^{\gamma}$ can affect the type of scattering, as discussed in Ref.~\onlinecite{huang2013stability}, and so in this general formulation we leave it unconstrained.

%#################################
%#################################

\subsection{Surface density of states}
\label{sec:LDOS}

The surface local density of states (LDOS) develops pronounced spatial inhomogeneities due to the impurities (Friedel oscillations, as discussed for WSMs in Ref.~\onlinecite{hosur2012friedel}). The total (spin-summed) surface LDOS at site $\textbf{r}_i$ relative to that of the clean system is given by,
\begin{equation}
\label{eq:LDOS_full}
\Delta\rho(\textbf{r}_i,\omega) = -\frac{1}{\pi} \text{Im Tr}~ \Delta \boldsymbol{G}^{0+}_{0+}(\textbf{r}_i,\textbf{r}_i,\omega)  \;,
\end{equation}
in terms of the surface Green function difference $\Delta \boldsymbol{G}^{0+}_{0+}(\textbf{r}_i,\textbf{r}_i,\omega) = \left [\boldsymbol{\mathcal{G}}^{0+}_{0+}(\textbf{r}_i,\textbf{r}_i,\omega)-\boldsymbol{G}^{0+}_{0+}(\textbf{r}_i,\textbf{r}_i,\omega) \right ]$, where $[\boldsymbol{G}^{j\tau_z}_{j'\tau_z'}(\textbf{r}_a,\textbf{r}_b,\omega)]_{\sigma\sigma'}\equiv \langle\langle \ann{c}{\textbf{r}_a,j\tau_z\sigma} ; \cre{c}{\textbf{r}_b,j'\tau_z'\sigma'} \rangle\rangle_{\omega}^0$ is a local real-space Green function for the clean system, while 
$[\boldsymbol{\mathcal{G}}^{j\tau_z}_{j'\tau_z'}(\textbf{r}_a,\textbf{r}_b,\omega)]_{\sigma\sigma'}$ is the corresponding full Green function, defined in the presence of the impurities.

The 2d Fourier transformation, Eq.~\ref{eq:rsop}, yields a surface momentum-space representation,
\begin{equation}
\label{eq:deltaG}
\Delta \boldsymbol{G}^{0+}_{0+}(\textbf{r}_i,\textbf{r}_i,\omega) = \int \frac{d^2 \kpar d^2 \kpar'}{(2\pi)^2}~\text{e}^{\text{i}\textbf{r}_i \cdot (\kpar'-\kpar)}~ \Delta \boldsymbol{G}^{0+}_{0+}(\kpar,\kpar',\omega) \;.
\end{equation}
The momentum-resolved surface density of states of the clean system is given by,
\begin{eqnarray}
\label{eq:dos_k}
\rho^0(\kpar,\omega)=-\tfrac{1}{\pi}\text{Im Tr}~\boldsymbol{G}^{0+}_{0+}(\kpar,\omega)\;.
\end{eqnarray}

%#################################
%#################################

\subsection{Surface t-matrix equation for single or multiple impurities}
\label{sec:tm}

Surface quasiparticles of the clean system with well-defined momentum $\kpar$ are scattered by impurities in disordered systems. This effect is described exactly by the scattering t-matrix equation. Here we formulate the t-matrix in terms of the scattering of \emph{surface} quasiparticles as required for QPI,\cite{derry2015quasiparticle} and generalize to the case of many impurities. One must retain the $2\times 2$ matrix spin-space structure inherent to the description of Dirac/Weyl systems. The surface t-matrix equation reads,
\begin{equation}
\label{eq:tm_def}
\Delta \boldsymbol{G}^{0+}_{0+}(\kpar,\kpar',\omega) = \boldsymbol{G}^{0+}_{0+}(\kpar,\omega) \boldsymbol{T}(\kpar,\kpar',\omega) \boldsymbol{G}^{0+}_{0+}(\kpar',\omega) \;,
\end{equation}
where $\boldsymbol{G}^{0+}_{0+}(\kpar,\omega)$ is obtained from Eq.~\ref{eq:GFsurf}, and $\boldsymbol{T}(\kpar,\kpar',\omega)$ is the t-matrix itself. For multiple generalized potential scattering impurities described by Eq.~\ref{eq:Himp}, the t-matrix can be expressed in terms of an infinite series of scattering events,
\begin{widetext}
\begin{equation}
\label{eq:tm_series}
\boldsymbol{T}(\kpar,\kpar',\omega)=\frac{1}{(2\pi)^2}\sum_{\gamma_0}\left [ \boldsymbol{V}^{\gamma_0}_{\kpar\kpar'} + 
\frac{1}{(2\pi)^2}\sum_{\gamma_1,\textbf{k}_1}\left [ \boldsymbol{V}^{\gamma_1}_{\kpar\textbf{k}_1} + 
\frac{1}{(2\pi)^2}\sum_{\gamma_2,\textbf{k}_2}\left [ \boldsymbol{V}^{\gamma_2}_{\kpar\textbf{k}_2} + ...
\right ] \boldsymbol{G}^{0+}_{0+}(\textbf{k}_2,\omega) \boldsymbol{V}^{\gamma_1}_{\textbf{k}_2\textbf{k}_1} 
\right ] \boldsymbol{G}^{0+}_{0+}(\textbf{k}_1,\omega) \boldsymbol{V}^{\gamma_0}_{\textbf{k}_1\kpar'} 
\right ] \;,
\end{equation}
\end{widetext}
where $\gamma_i$ are impurity labels, and $\boldsymbol{V}^{\gamma_i}_{\textbf{k}_a\textbf{k}_b} = \boldsymbol{V}^{\gamma_i} \text{e}^{\text{i} \textbf{r}_{\gamma_i}\cdot (\textbf{k}_a-\textbf{k}_b)}$. The infinite matrix series in Eq.~\ref{eq:tm_series} can be summed to give a compact exact expression,
\begin{equation}
\label{eq:tm}
\boldsymbol{T}(\kpar,\kpar',\omega)= \frac{1}{(2\pi)^2} \sum_{\gamma,\gamma'} \text{e}^{\text{i} (\textbf{r}_{\gamma}\cdot \kpar' - \textbf{r}_{\gamma'}\cdot \kpar)} \times \left [ \varmathbb{T}(\omega) \right ]_{\gamma\gamma'} \;,
\end{equation}
where $\varmathbb{T}(\omega)$ is an $N\times N$ matrix for an $N$-impurity system, with elements $[\varmathbb{T}(\omega)]_{\gamma\gamma'}$ that are themselves $2\times 2$ matrices. It is given by,
\begin{equation}
\label{eq:T_def}
\varmathbb{T}(\omega)= \varmathbb{V}\left [ \boldsymbol{I}-\varmathbb{G}(\omega)\varmathbb{V} \right ]^{-1} \;,
\end{equation}
where $[\varmathbb{V}]_{\gamma\gamma'}=\boldsymbol{V}^{\gamma}\delta_{\gamma\gamma'}$ and $[\varmathbb{G}(\omega)]_{\gamma\gamma'}=\boldsymbol{G}^{0+}_{0+}(\textbf{r}_{\gamma},\textbf{r}_{\gamma'},\omega)$ for impurities $\gamma$ and $\gamma'$.

In the dilute impurity limit with just a single impurity located at $\textbf{r}_0$ on the surface, Eq.~\ref{eq:tm} reduces to,
\begin{equation}
\label{eq:tm_1imp}
\boldsymbol{T}(\kpar,\kpar',\omega)= \frac{1}{(2\pi)^2} \boldsymbol{V}^0 \left [ \boldsymbol{I}-\boldsymbol{G}^{0+}_{0+}(\textbf{r}_0,\textbf{r}_0,\omega) \boldsymbol{V}^0 \right ]^{-1} \;.
\end{equation}

In the weak-scattering `Born' limit, one approximates the full complex and dynamical t-matrix by the real static quantity $\boldsymbol{T}(\kpar,\kpar',\omega) \approx \frac{1}{(2\pi)^2} \boldsymbol{V}^0$. In the following numerical calculations, there is no need to resort to the Born approximation, and the full expression for a single impurity, Eq.~\ref{eq:tm_1imp}, is used. For concreteness, we now take $\boldsymbol{V}^0=V\boldsymbol{I}$, with $V=0.01$.

%#################################
%#################################

\subsection{Magnetic (Kondo) impurities}
\label{sec:kondo}

The physics of (dynamic) magnetic impurities, such as transition metal adatoms, is of course very rich due to strong electron correlations and the Kondo effect.\cite{hewson1997kondo,bulla2008numerical} In 3D Weyl systems, various unusual Kondo variants can occur, as discussed recently in Ref.~\onlinecite{mitchell2015kondo}. The t-matrix then develops nontrivial dynamics, requiring a sophisticated many-body treatment. In particular, Kondo-enhanced spin-flip scattering can lead to low-energy resonances, characterized by a t-matrix with a large imaginary part (the Born approximation is therefore totally inapplicable). However, if the t-matrix is obtained for a given Kondo system (e.g.\ from a numerical renormalization group calculation\cite{mitchell2015kondo}), it can simply be used instead of Eq.~\ref{eq:tm_1imp} in the following.

%#################################
%#################################

\section{Quasiparticle Interference (QPI)}
\label{sec:QPI}
QPI is obtained experimentally via FT-STS.\cite{hoffman2002imaging,zheng2015atomic} It involves local measurements on the surface using STS to produce the real-space LDOS map $\rho(\textbf{r}_i,\omega)$ at a given tip-sample bias voltage $\propto \omega$. Due to the presence of impurities, this LDOS map shows pronounced spatial inhomogeneities. Its 2D Fourier transform yields the QPI, which characterizes the scattering of surface quasiparticles of the WSM material due to the impurities. The preferred QPI scattering vectors reveal the electronic structure of the Weyl system --- and as we show in the following, the QPI also reveals its topological structures. 
The QPI is defined as,
\begin{equation}
\label{eq:QPI_RSdef}
\Delta\rho(\bold{q},\omega) = \sum_{i} \text{e}^{-\text{i}\textbf{q}\cdot \textbf{r}_i}~\Delta\rho(\textbf{r}_i,\omega) \;,
\end{equation}
in terms of the LDOS difference, Eq.~\ref{eq:LDOS_full}. The QPI can be alternatively expressed in terms of momentum-space surface Green functions by using Eqs.~\ref{eq:LDOS_full} and \ref{eq:deltaG} in Eq.~\ref{eq:QPI_RSdef}. For the WSM, it can be shown that,
\begin{equation}
\label{eq:QPI_Qdef}
\Delta\rho(\bold{q},\omega) = -\frac{1}{2\pi \text{i}} \text{Tr}~\left [ Q(\bold q,\omega) - Q(-\bold q,\omega)^* \right ] \;,
\end{equation}
with,
\begin{equation}
\label{eq:Q_def}
\begin{split}
Q(\bold q,\omega) &= \text{Tr}\int d^2 \kpar ~\Delta \boldsymbol{G}^{0+}_{0+}(\kpar,\kpar-\bold q,\omega) \;, \\
 &=  \text{Tr}\int d^2 \kpar ~\boldsymbol{G}^{0+}_{0+}(\kpar,\omega) \boldsymbol{T}(\kpar,\kpar-\bold q) \boldsymbol{G}^{0+}_{0+}(\kpar-\bold q,\omega) \;,
\end{split}
\end{equation}
where the second line follows from the definition of the t-matrix in Eq.~\ref{eq:tm_def}. 

Although FT-STS and QPI probe the surface, it is important that the Green's functions $\boldsymbol{G}^{0+}_{0+}(\kpar,\omega)$ that enter Eq.~\ref{eq:Q_def} are defined for the full 3D system. Unlike 3D topological insulators, where the Dirac cones live on the 2D surface and an effective surface continuum theory can be constructed (although a two dimensional lattice formulation is not possible), there is for instance no simple effective surface theory for the Fermi arc since it can be viewed as one Fermi surface split over the upper and lower surfaces of the material. Importantly, surface Green's functions for a 3D system contain information about electronic propagation from the surface, into the bulk, and back to the surface. The fact that surface quasiparticles in WSMs are `dephased' by coupling to the bulk allows for richer QPI structures, such as open Fermi arcs, that cannot arise in pure 2D systems.

Note also that the matrix structure of Eq.~\ref{eq:Q_def} implies that both spin-diagonal and spin-off-diagonal elements of the $2\times 2$ Green's function matrix $\boldsymbol{G}^{0+}_{0+}(\kpar,\omega)$ enter. Furthermore, both real and imaginary parts of the Green's functions are important in obtaining the correct QPI. Clearly, more information is contained in the QPI than simply the total surface density of states $\rho^0(\kpar,\omega)$, as extracted from ARPES experiments. 
As highlighted in Ref.~\onlinecite{derry2015quasiparticle}, the QPI cannot be understood in terms of the joint density of states except in the simplest of cases; it is certainly not applicable to dynamical multiband systems and topological materials such as WSMs -- see Sec.~\ref{sec:JDOS}.

In this paper, we obtain the exact QPI from Eqs.~\ref{eq:QPI_Qdef}, \ref{eq:Q_def}, using the surface Green functions $\boldsymbol{G}^{0+}_{0+}(\kpar,\omega)$ from Eq.~\ref{eq:GFsurf}, and the t-matrix $\boldsymbol{T}(\kpar,\kpar')$ due to a single potential scattering impurity from Eq.~\ref{eq:tm_1imp}. Throughout, we show the QPI as a color plot on a scale relative to the most intense scattering vector. Generally, the overall intensity increases with scanning energy, and is proportional to the static impurity scattering potential, $V$.

Scattering from multiple (uncorrelated) impurities leads to an overlaid moir\'{e} pattern in the QPI (see e.g.\ the explicit calculations of Ref.~\onlinecite{mitchell2015multiple}). When the real-space surface region probed by STS is large enough that many randomly distributed impurities contribute to the measured scattering, these additional QPI structures average out and yield a good approximation to the pristine single-impurity result considered here.

%#################################
%#################################

\section{Weyl nodes and Dirac cones in QPI}
\label{sec:Dirac}

%################
\begin{figure}[t]
\begin{center}
\includegraphics[width=90mm]{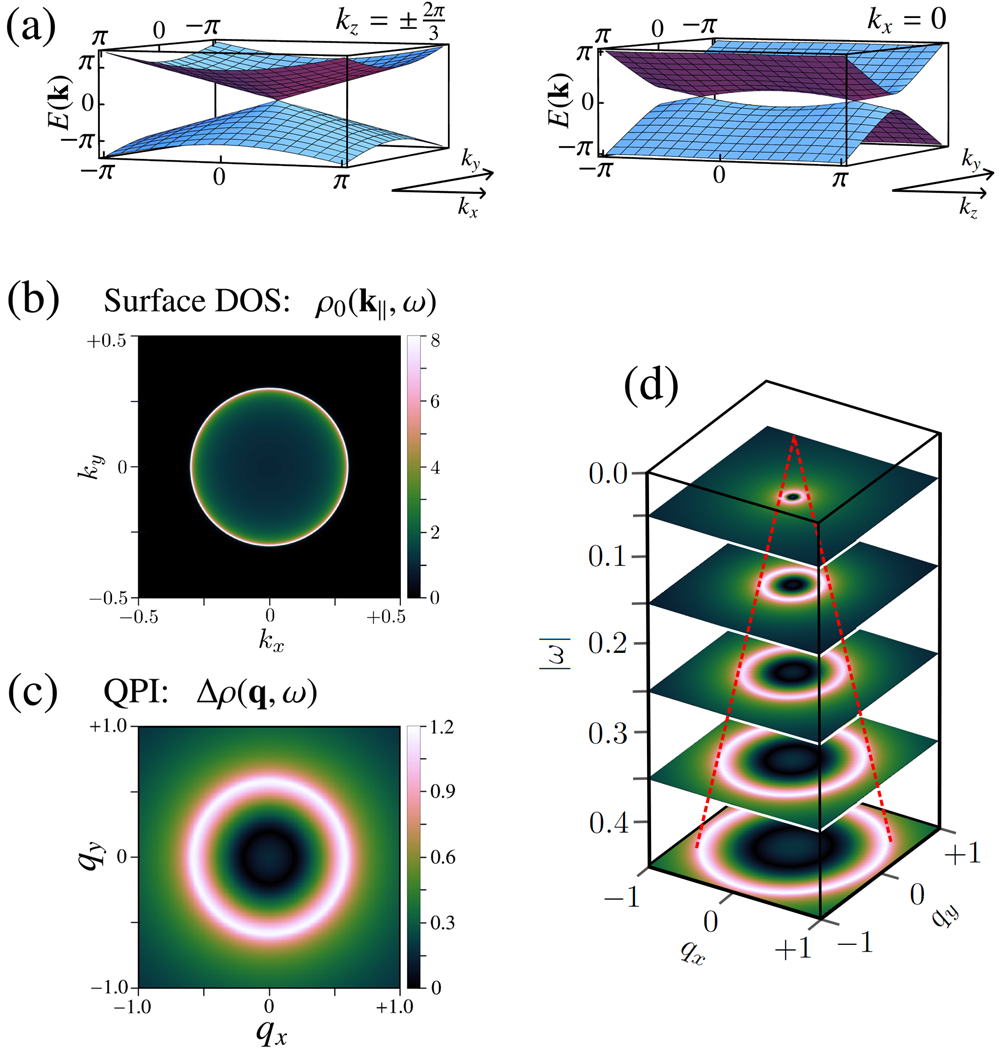}
\caption{\label{fig:dirac}
Weyl semimetal realized by breaking TRS. Plotted for $\Delta_{T}(\kpar)=\Delta_{N}(\kpar)\equiv\Delta$ and $\delta(\kpar)=0$, with $m_z=\Delta=\tfrac{1}{2}$. (a) Bulk band structure, showing a single pair of separated 3D Weyl nodes at $\textbf{k}\equiv (k_x,k_y,k_z)=(0,0,\pm\tfrac{2\pi}{3})$. (b) Momentum-resolved DOS $\rho_0(\kpar,\omega)$ at the surface in the $k_x k_y$ plane, at bias voltage (scanning energy) $\omega=0.3$. (c) Corresponding QPI, $\Delta \rho(\textbf{q},\omega)$. (d) Stack plot showing the Dirac cone structure in the QPI mapped out on increasing bias voltage. Fermi arcs are not observed in either the surface DOS or QPI because the inter-node vector is perpendicular to the surface.
}
\end{center}
\end{figure}
%################

We consider now a Weyl system featuring a single pair of nodes in the 3D bulk. For the system described by Eqs.~\ref{eq:H}--\ref{eq:TIsurface}, the TRS-breaking perturbation $m_z$ can render the Weyl nodes nondegenerate, separating them in momentum space along $k_z$. 

We take the inversion symmetric case $\delta(\kpar)=0$, and assume for simplicity rotational symmetry with $\Delta_N(\kpar)=\Delta_T(\kpar)\equiv \Delta$. For $\Delta=m_z=\tfrac{1}{2}$, the Weyl nodes are located at $\textbf{k}=(0,0,\pm\tfrac{2\pi}{3})$, as shown from the bulk band structure in Fig.~\ref{fig:dirac} (a). In this case, analysis of Eq.~\ref{eq:GFsurf} yields a simple exact expression for the surface Green's functions,
\begin{equation}
\label{eq:exactG_mz}
\begin{split}
G^{0+\uparrow}_{0+\uparrow}(\kpar,z) =& \frac{1}{8 m_z \Delta^2 (k^2-z^2)}\Bigg [ 4 m_z (z-m_z)(k^2-z^2) \\
&- \left(k^2-(z-m_z)^2+4\Delta^2-\phi\right) \\
&\times \sqrt{2(k^2-z^2)(k^2-z^2-m_z^2+4\Delta^2+\phi)} \Bigg ]\;,
\end{split}
\end{equation}
where $z=\omega+\text{i}0^{+}$, $k=v_F|\kpar |$ and $\phi^2 = (k^2-z^2+(m_z-2\Delta)^2)(k^2-z^2+(m_z+2\Delta)^2)$. Note that by symmetry  $G^{0+\uparrow}_{0+\uparrow}(\kpar,\omega+\text{i}0^{+})=G^{0+\downarrow}_{0+\downarrow}(\kpar,-\omega+\text{i}0^{+})$. Expressions for the off-diagonal elements of $\boldsymbol{G}^{0+}_{0+}(\kpar,z)$ can also be simply obtained. 

The singular structure of the surface states is already apparent from the denominator of Eq.~\ref{eq:exactG_mz}. The classic Dirac cone structure appears in the surface Green functions, with divergent rings $v_F |\kpar|=|\omega|$ at a given energy (surface-tip bias) $\omega$, and strictly excluded spectral weight for all $v_F |\kpar|>|\omega|$. The momentum-resolved surface density of states for the clean system, $\rho^0(\kpar,\omega)$, is plotted in Fig.~\ref{fig:dirac} (b) for bias voltage $\omega=0.3$. No Fermi arcs are observed in this system because the inter-node vector along $\hat{z}$ is perpendicular to the surface in the $xy$ plane.

Fig.~\ref{fig:dirac} (c) shows the corresponding QPI at the same bias voltage (but note the doubled axis scales). A cross-section through the Dirac cone is observed, with the ring $v_F|\textbf{q}|=2|\omega|$ corresponding to the most intense scattering. However, unlike the surface density of states in (b), there is finite scattering for all $\textbf{q}$ (the difference is due to the  matrix structure of Eq.~\ref{eq:Q_def}, which also involves complex Green's functions rather than spectral densities).

The full surface-projected Dirac cone structure associated with bulk Weyl nodes can be mapped by scanning the bias voltage, as demonstrated in Fig.~\ref{fig:dirac} (d). FT-STS can therefore be considered as a complementary probe to ARPES, where similar structures have been observed in 3D Dirac\cite{jeon2014landau,liu2014stable} and Weyl\cite{Lv2015Experimental,xu2015discovery,xu2015discovery2,yang2015weyl,lv2015observation,xu2015observation} systems.

%#################################
%#################################

%################
\begin{figure}[t]
\begin{center}
\includegraphics[width=70mm]{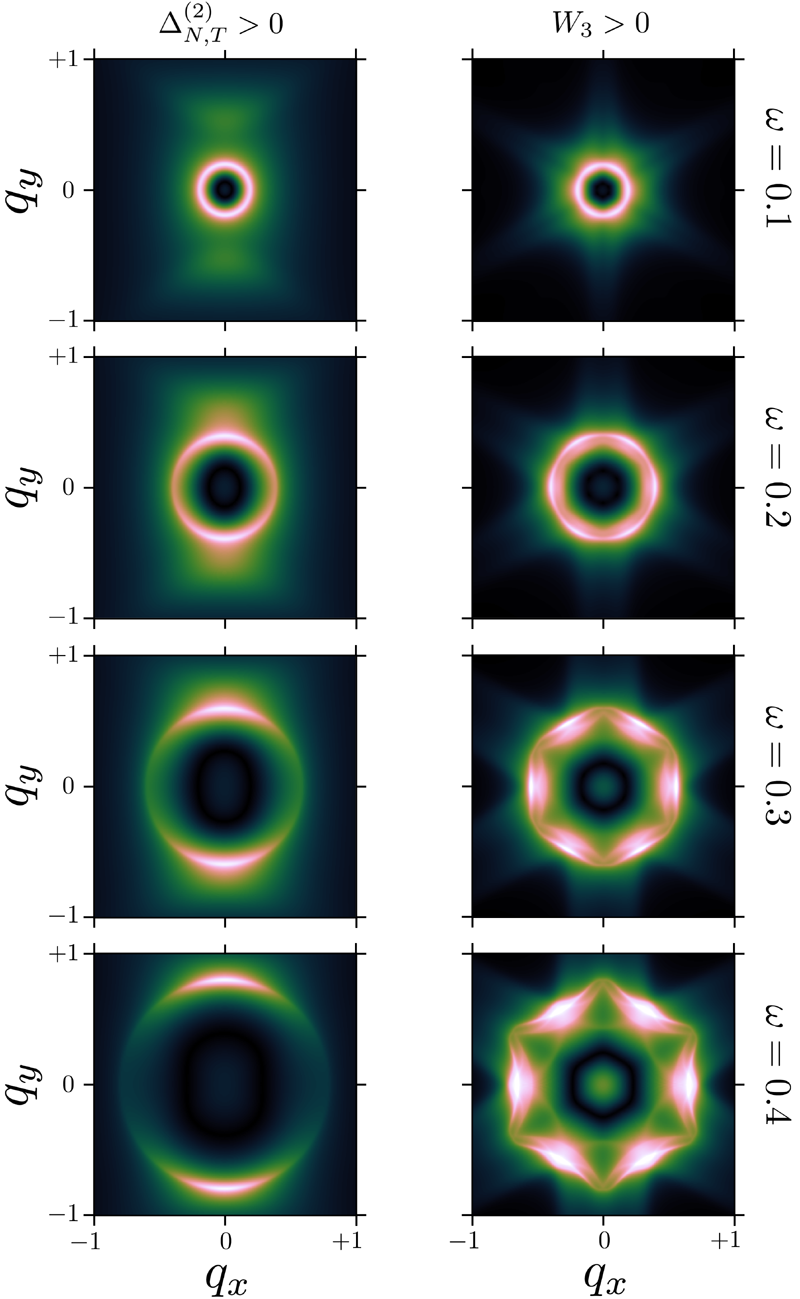}
\caption{\label{fig:warp}
Effect of crystal warping on the Dirac cone structure of QPI in Weyl semimetals. Plotted for the TRS breaking case with $m_z=0.1 $ and $\Delta_{N,T}^{(0)}=\tfrac{1}{2} $, including tunneling anisotropy $\Delta_{N,T}^{(2)}= 1$ (left panels) and cubic warping $W_3=5 $ (right panels) for increasing bias voltages $\omega = 0.1, 0.2, 0.3, 0.4$ (top to bottom). The rotational symmetry of the Dirac cone seen at low energies is strongly modified at higher energies. The dispersive arc-like features are not Fermi arcs.
}
\end{center}
\end{figure}
%################

\section{Crystal Warping of Dirac cones in QPI}
\label{sec:warp}

At low energies, bulk Weyl nodes are characterized by the rotational symmetry of Eq.~\ref{eq:DiracSM} (an anisotropic Fermi velocity $v_F\rightarrow \vec{v}_F$ leads only to a simple rescaling). Crystal warping, due to the underlying lattice structure of the material, does not destroy the Weyl nodes by opening a gap, but it does spoil the rotational symmetry of the Dirac cones at higher energies. 
Experiments on the 3D Dirac semimetal system Cd$_3$As$_2$ in Ref.~\onlinecite{jeon2014landau} appear to show crystal warping effects in the measured QPI. Such effects might similarly be important in interpreting QPI patterns for Weyl systems away from the Fermi energy.\cite{zheng2015atomic}

As discussed in Sec.~\ref{sec:crystal_warp}, a two-fold symmetry of the lattice in the $xy$ plane can be phenomenologically encoded in the effective Hamiltonian via finite $\Delta_{N,T}^{(2)}$; while cubic warping implies finite $W_3$. The Weyl nodes are topologically robust to such perturbations.
The effect on QPI of each is illustrated in Fig.~\ref{fig:warp} for the TRS-broken Weyl system of Fig.~\ref{fig:dirac} ($m_z>0$). The surface Green's functions are only described by Eq.~\ref{eq:exactG_mz} in the low-bias limit; the full QPI was therefore computed numerically in these cases. 
$\Delta_{N,T}^{(2)}>0$ is shown in the left panels, while $W_3>0$ is shown in the right panels, with increasing bias voltage from top to bottom.  At low energies $\omega=0.1$, the cross-section of the Dirac cone is essentially circular in both cases. However, at higher energies the symmetry of the underlying lattice reveals itself through the preferred QPI scattering vectors.

%################
\sidecaptionvpos{figure}{}
\begin{SCfigure*}
\includegraphics[width=120mm]{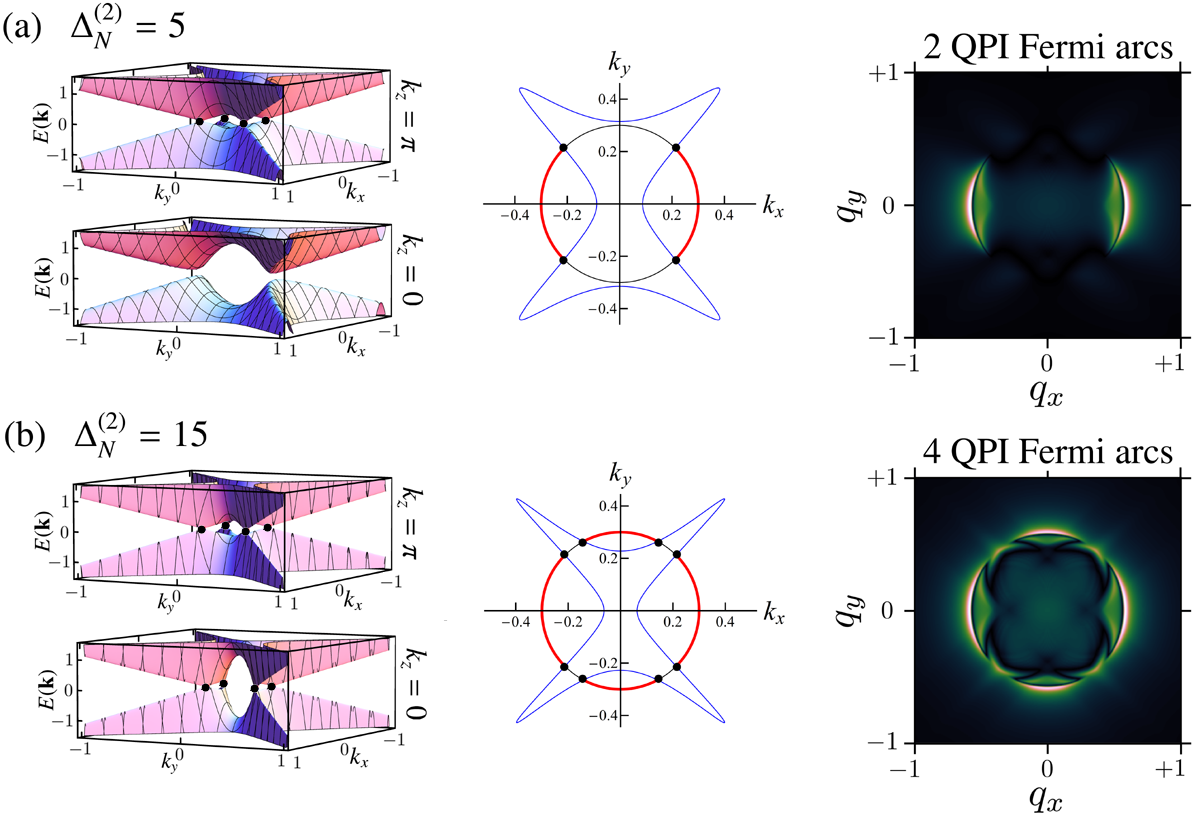}
\caption{\label{fig:arc}
QPI Fermi arcs in time reversal invariant Weyl semimetals. Shown for $|\vec{m}|=0$, $\delta(\kpar)\equiv \delta=0.3$ and  $\Delta_{N,T}^{(0)}=\tfrac{1}{2} $. \emph{Left:} Bulk band structure in the $k_x k_y$ plane parallel to the surface at $k_z=\pi$ and $0$. Black points indicate Weyl points. \emph{Center:} Singular structures in the surface Green's functions at the Fermi energy, Eq.~\ref{eq:singular} (blue lines). Black lines show $v_F \textbf{k}_{\parallel}=\delta$, the line node ring in the isotropic case. Black dots show the surface projection of the Weyl points, while the overlaid thick red lines indicate Fermi arcs. \emph{Right:} 
QPI plotted in the $q_x q_y$ plane at very low bias voltage $\omega=0.01$. 
(a) $\Delta_{N}^{(2)}=5 $: two pairs of Weyl nodes at $k_z=\pi$, manifest on the surface in QPI as two Fermi arcs.
(b) $\Delta_{N}^{(2)}=15 $: two pairs of Weyl nodes at $k_z=\pi$ and another two pairs at $k_z=0$, producing four QPI Fermi arcs. 
}
\end{SCfigure*}
%################

Note that Fermi arcs are not seen in this case because of the relative orientation of the surface to the inter-node vector: at the Fermi energy, scattering of surface quasiparticles is confined to the point $\textbf{q}=\textbf{0}$. By contrast, dispersive arc-like features can be observed in QPI at higher energies due to crystal warping (e.g.\ finite $\Delta_{N,T}^{(2)}>0$ -- see bottom left panels of Fig.~\ref{fig:warp}). These are not topological features, since they disappear at low energies.

%#################################
%#################################

\section{Fermi arcs in QPI}
\label{sec:FA}

Arguably the most distinctive feature of WSMs is the existence of Fermi arcs in the surface DOS,\cite{wan2011topological} due to topologically-protected zero-energy states connecting bulk Weyl nodes of opposite chirality, projected onto the surface. Fermi arcs have recently been observed experimentally in the monopnictide Weyl materials using ARPES to probe the momentum-resolved surface DOS.\cite{Lv2015Experimental,xu2015discovery,xu2015discovery2,yang2015weyl,lv2015observation,xu2015observation} 

We now discuss how impurity-induced quasiparticle scattering from these surface states at the Fermi energy also produces intense and characteristic signatures in QPI.

Since the present experimental WSM candidates break IS rather than TRS, we now examine IS-broken but TR-invariant Weyl systems, characterized at low-energies by finite $\delta(\kpar)\equiv\delta$. We focus on experimentally-relevant situations where the system supports several pairs of Weyl nodes. The relative contribution from inter-arc and intra-arc scattering can then be assessed.

Specifically, we take $\Delta_{N}^{(0)}=\Delta_{T}^{(0)}=\tfrac{1}{2}$ and $|\vec{m}|=0$. In the rotationally-symmetric case with no crystal warping $\Delta_{N,T}^{(n>0)}=0$, the system supports a bulk \emph{line} node in the $k_x k_y$ plane parallel to the surface. Surface Green's functions exhibit a singular structure in a ring with  $v_F |\kpar|=|\omega-\delta|$. The surface DOS, related to the imaginary part of the surface Green's functions, diverges on approaching this ring. The real part of the surface Green's function has a definite sign in the region enclosed by the ring (the complementary region is of opposite sign).

%################
\begin{figure*}[t]
\begin{center}
\includegraphics[width=160mm]{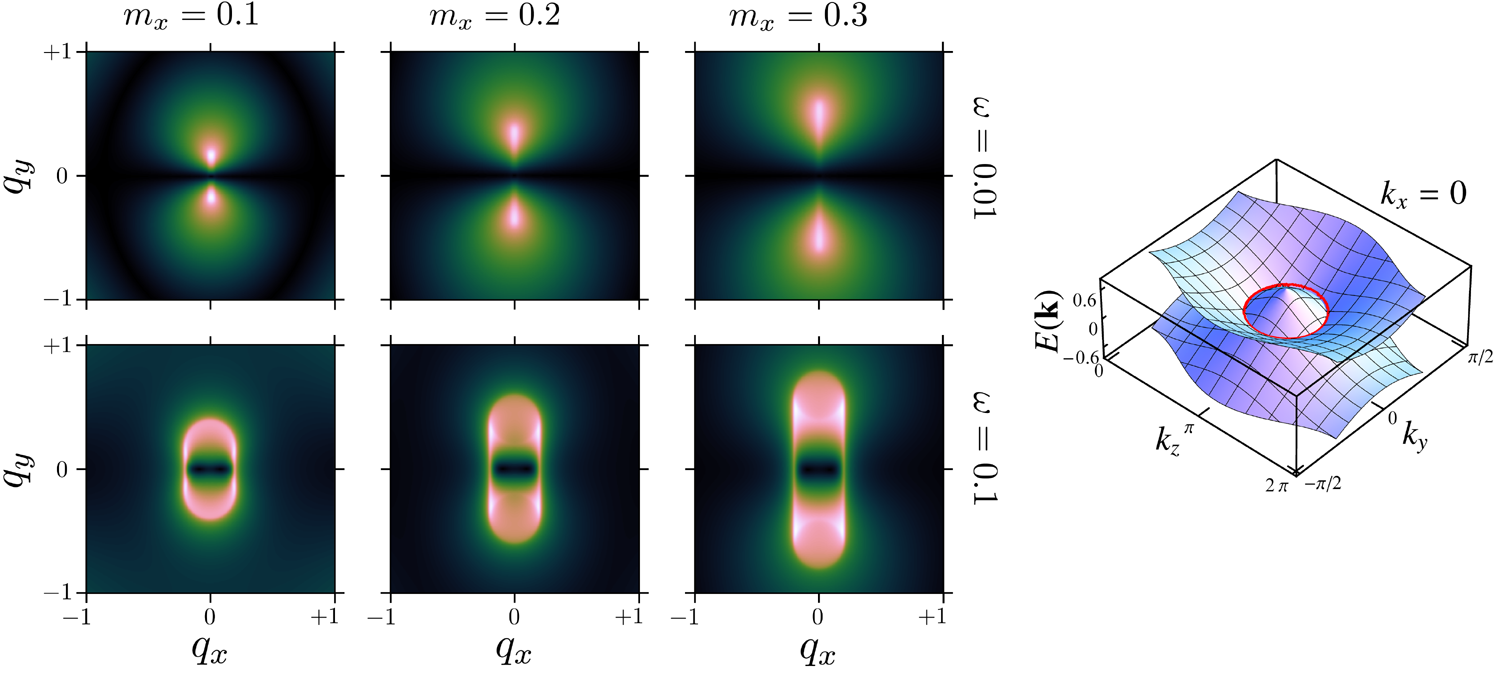}
\caption{\label{fig:line}
QPI for a line-node semimetal. Plotted for the TRS-broken case $m_x>0$, with $\Delta_{N,T}^{(0)}=\tfrac{1}{2} $ and $\delta(\kpar)=0$. The band structure shown on the right is characterized by a line node ring (indicated in red) at $k_x=0$ in the $k_y k_z$ plane perpendicular to the surface. The surface projection in QPI is shown for  $m_x=0.1, 0.2, 0.3$ (left to right), at scanning energy $\omega=0.01$ (upper panels) and $0.1$ (lower panels).
}
\end{center}
\end{figure*}
%################

As noted in Ref.~\onlinecite{halasz2012time}, such a line node is delicate and can be destroyed when the continuous rotational symmetry is reduced to a discrete point symmetry, realizing chiral pairs of Weyl nodes. This is naturally achieved by tunneling anisotropy, as might be expected in real crystals, and we now take $\Delta_{N}^{(2)}>0$. In Fig.~\ref{fig:arc}(a) we consider $\Delta_{N}^{(2)}=5$, while stronger anisotropy $\Delta_{N}^{(2)}=15$ is used for (b). The bulk band structure in the $k_x k_y$ plane at $k_z=\pi$ and $0$ is shown on the left in each case. For $\Delta_{N}^{(2)}=5$ in (a), two pairs of Weyl nodes are seen at $k_z=\pi$. On increasing the anisotropy, we find that two additional pairs of degenerate (Dirac) nodes appear at $k_z=0$ when $\delta^2=\Delta^{(0)}/\Delta_N^{(2)}$. The degeneracy is lifted as the Weyl nodes are split in momentum space on further increasing the anisotropy, as shown in Fig.~\ref{fig:arc}(b) for $\Delta_{N}^{(2)}=15$. The Weyl nodes are indicated with black points for clarity.

The exact surface Green's functions, obtained from solutions of Eq.~\ref{eq:GFsurf}, can again be found analytically in closed form. In the present case they are rather complicated and so we do not give them in full. At the Fermi energy $\omega=0$, however, analysis reveals a singular structure along the line,
\begin{equation}
\label{eq:singular}
\begin{split}
&8\left(v_F\Delta_N^{(2)}\right)^2\left[k_x^2 - k_y^2\right]
+ 4 \Delta_N^{(2)}\Delta^{(0)}-1 \\ 
&=\pm\sqrt{1-8 \Delta_N^{(2)} \Delta^{(0)}+16 \left(\Delta_N^{(2)}\right)^2 \left[\left(\Delta^{(0)}\right)^2 - 2v_F^2 k_x^2 +\delta^2\right]} \;,
\end{split}
\end{equation}
shown as the blue lines in the center column panels of Fig.~\ref{fig:arc}. This singular line is identified from zeros of the denominator of the surface Green's functions. The real part of the Green's function in the region enclosed by this line again has a definite sign; the sign changes as the blue line is crossed. 
For $\delta^2<\Delta^{(0)}/\Delta_N^{(2)}$ in (a), the singular line crosses $v_F \kpar=\delta$ at four points; two pairs of Weyl points arise at these intersections (marked as black points). This is consistent with the bulk band structure calculations shown in the left panel, where the Weyl points are found at $k_z=\pi$. When $\delta^2>\Delta^{(0)}/\Delta_N^{(2)}$, as in Fig.~\ref{fig:arc}(b), two further pairs of Weyl points appear (from the bulk band structure in the corresponding left panel, the new Weyl points can be associated with $k_z=0$).

Fermi arcs are found to exist in regions with a definite parity of the real part of the Green's functions, and are therefore terminated on intersection with the blue line, at surface projections of the Weyl points. The Fermi arcs connect chiral pairs of Weyl points, and are indicated in the center panels of Fig.~\ref{fig:arc} by the red lines. The appearance of the second pair of Fermi arcs on increasing the anisotropy $\Delta_N^{(2)}$ therefore signals the topological change in the system as additional Weyl node pairs are created.

The right panels of Fig.~\ref{fig:arc} show the calculated QPI for these cases, close to the Fermi energy at $\omega=0.01$. QPI Fermi arcs of intense scattering are observed, connecting projections of the Weyl nodes to the surface. As the anisotropy is tuned through $\delta^2=\Delta^{(0)}/\Delta_N^{(2)}$, two additional QPI Fermi arcs appear, connecting the new Weyl points. 

We note that there is intense \emph{inter}-arc scattering here (producing QPI Fermi arcs at doubled $\textbf{q}$-vectors), but comparatively very weak \emph{intra}-arc scattering (which might be expected to produce QPI features around $\textbf{q}=\textbf{0}$). This `extinction' of quasiparticle scattering can be attributed to quantum interference effects; see also Sec.~\ref{sec:JDOS} below.

%#################################
%#################################

\section{Line nodes in QPI}
\label{sec:LN}

Finally we consider the case of line node semimetals, realized here by breaking TRS through finite $m_x>0$. For $\delta(\kpar)=0$ and $\Delta_{N,T}(\kpar)\equiv \Delta$, the line node is a ring in the $k_y k_z$ plane centered on $\textbf{k}=\textbf{0}$. This is shown as the red line at zero energy in the band structure diagram on the right of Fig.~\ref{fig:line}. The plane containing the line node is perpendicular to the surface. 

To understand the signatures of this kind of bulk topological structure on the surface in QPI, we first consider the surface Green's functions from Eq.~\ref{eq:GFsurf}. Analysis shows that these Green's functions contain singular lines that satisfy,
\begin{equation}
\label{eq:line}
\left(v_F^2 k_x^2-\omega^2 \right)\left(v_F^2 k_x^2+(v_F k_y\pm m_x)^2-\omega^2\right)=0 \;.
\end{equation}
Interestingly, there are structures resembling Dirac cones, centered on $v_F \kpar=(0,\pm m_x)$, and connected by lines at $v_F k_x=\pm \omega$. These lead to lines of intense scattering in QPI.

The full numerically-calculated QPI is shown in Fig.~\ref{fig:line} for $m_x=0.1, 0.2, 0.3$ (left to right panels) at bias voltages $\omega=0.01$ and $0.1$ (upper and lower panels, respectively). 
At low energies, the surface projection of the line node leads to a line in QPI connecting $v_F \textbf{q}=(0,\pm 2 m_x)$. However, quasiparticle scattering is only intense at the terminal points -- see upper panels. This distinguishes them from the QPI Fermi arcs arising for separated Weyl nodes. At higher energies (lower panels), tube-like projections of the bulk structure appear in the surface QPI. These line-node signatures can be distinguished from those of Dirac cones since scattering is intense for all $[v_F^2 k_x^2+(v_F k_y\pm m_x)^2]\le \omega^2$.

%%%%%%%%%%%%%%%%%%%%%%%%%%%%%%%%%%%%%%%%%%%%%%%%%%%%%%%%%%%%%%%%%%%%%%%
%%%%%%%%%%%%%%%%%%%%%%%%%%%%%%%%%%%%%%%%%%%%%%%%%%%%%%%%%%%%%%%%%%%%%%%

\section{Failure of the joint-density-of-states approach and ``extinction''}
\label{sec:JDOS}

QPI data is often interpreted in terms of the phenomenological JDOS approach, which can provide an intuitive rationalization of the preferred QPI scattering vectors. The JDOS is defined as,
\begin{equation}
\label{eq:jdos}
J(\textbf{q},\omega)=\int d^2 \kpar ~\rho^0(\textbf{k}_{\parallel},\omega)\rho^0(\textbf{k}_{\parallel}+\textbf{q},\omega) \;,
\end{equation}
in terms of the momentum-resolved surface DOS of the clean system, given by Eq.~\ref{eq:dos_k}. 

However, we stress that FT-STS experiments measure the QPI, not the JDOS. The JDOS, Eq.~\ref{eq:jdos}, cannot be derived from the full QPI, Eq.~\ref{eq:QPI_Qdef} in any limit. In the special case of a single static impurity in a one-band, particle-hole symmetric host, the JDOS and QPI are related by a Kramers-Kronig transformation and so do share some common features;\cite{derry2015quasiparticle} but this is not the case for general multiband problems, such as those of the WSMs. 

To fully capture quantum interference effects of different scattering pathways in the QPI, one must account for the \emph{phase} of the complex Green's functions in Eq.~\ref{eq:Q_def} (rather than using only the imaginary part, as in the JDOS). Furthermore, the matrix structure of Eq.~\ref{eq:Q_def} means that spin-off-diagonal scattering is included, whereas the trace in Eq.~\ref{eq:dos_k} neglects this information. Finally, we note that the T-matrix itself can be complex.

\begin{figure}[t]
\begin{center}
\includegraphics[width=80mm]{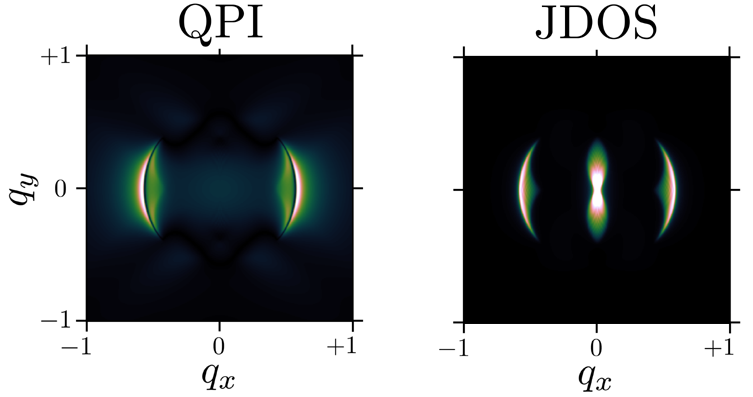}
\caption{\label{fig:jdos}
Comparison of the full QPI (left) and JDOS (right) for the system in Fig.~\ref{fig:arc}(a), which hosts two surface Fermi arcs. The JDOS was obtained using Eq.~\ref{eq:jdos}.
}
\end{center}
\end{figure}

We demonstrate explicitly the failure of the JDOS approach in Fig.~\ref{fig:jdos}, where we compare the full QPI to the JDOS, for a system with two Fermi arcs [using the same parameters as Fig.~\ref{fig:arc}(a)]. The JDOS correctly predicts the existence of QPI Fermi arcs, due to inter-arc scattering, but also spuriously predicts a figure-of-8 structure around $\textbf{q}=\textbf{0}$, attributable to intra-arc scattering. This feature is absent in the true QPI, and is therefore an example of `extinction' of quasiparticle scattering.

Recently, it was shown in Ref.~\onlinecite{kourtis2015universal} that such pinch-point structures appear ubiquitously in the JDOS for Weyl systems with Fermi arcs in their surface DOS. However, we point out that they may or may not appear in the measurable QPI, depending on quantum interference effects.

In general, the JDOS should not be expected to reproduce (even qualitatively) the QPI for such materials. We have also verified that the spin-dependent scattering probability (SSP)\cite{kourtis2015universal} similarly fails for this model.

%################
%################

%%%%%%%%%%%%%%%%%%%%%%%%%%%%%%%%%%%%%%%%%%%%%%%%%%%%%%%%%%%%%%%%%%%%%%%
%%%%%%%%%%%%%%%%%%%%%%%%%%%%%%%%%%%%%%%%%%%%%%%%%%%%%%%%%%%%%%%%%%%%%%%

\section{Conclusion}
\label{sec:conc}

Quasiparticle interference, obtained experimentally through FT-STS, is a powerful and sensitive tool for detecting, imaging, and distinguishing topological features in Weyl systems. Although FT-STS is a surface probe, the QPI reveals surface projections of nontrivial bulk topology. Furthermore, QPI offers simultaneous momentum and energy resolution, and contains more information on the band structure than is contained just in the density of states.

We presented a general framework for calculation of QPI in systems with an explicit surface, based on a Green's function formalism. The scattering problem due to a single impurity (dilute limit), or disorder from many impurities, is characterized in terms of the t-matrix. The approach goes beyond the `joint density of states' approximation, which cannot in general reproduce the complexities of the true QPI for multiband topological systems.

The QPI is shown to exhibit distinctive and characteristic features for WSMs, depending on the topology of the bulk, and the relative surface orientation. We studied a range of systems, including in particular a time-reversal invariant model with broken inversion symmetry, hosting several pairs of Weyl nodes. QPI Fermi arcs, resulting from intense inter-arc scattering, were found to appear in this case, although (in our model) there was extinction of intra-arc quasiparticle scattering due to quantum interference effects. We also showed how Dirac cone structures can be mapped out in QPI as a function of bias voltage; we studied the effect of crystal warping at higher energies, and investigated the possible signatures of more exotic line node WSMs.

FT-STS experiments should therefore provide valuable new insights into topological Weyl materials such as the monopnictides.

%%%%%%%%%%%%%%%%%%%%%%%%%%%%%%%%%%%%%%%%%%%%%%%%%%%%%%%%%%%%%%%%%%%%%%%
%%%%%%%%%%%%%%%%%%%%%%%%%%%%%%%%%%%%%%%%%%%%%%%%%%%%%%%%%%%%%%%%%%%%%%%

\acknowledgements
This work is part of the D-ITP consortium, a program of the Netherlands Organisation for Scientific Research (NWO) that is funded by the Dutch Ministry of Education, Culture and Science (OCW).

%%%%%%%%%%%%%%%%%%%%%%%%%%%%%%%%%%%%%%%%%%%%%%%%%%%%%%%%%%%%%%%%%%%%%%%
%%%%%%%%%%%%%%%%%%%%%%%%%%%%%%%%%%%%%%%%%%%%%%%%%%%%%%%%%%%%%%%%%%%%%%%

%\bibliography{refs}


%%%%%%%%%%%%%%%%%%%%%%%%%%%%%%%%%%%%%%%%%%%%%%%%%%%%%%%%%%%%%%%%%%%%%%%
%%%%%%%%%%%%%%%%%%%%%%%%%%%%%%%%%%%%%%%%%%%%%%%%%%%%%%%%%%%%%%%%%%%%%%%

\end{document}